\documentclass[reprint,superscriptaddress,floatfix,amsmath,amssymb,pra,aps,showpacs,twocolumn]{revtex4-2}
\usepackage{amsmath,amsfonts,amssymb,amsthm,graphics,graphicx,epsfig,bbm}
\usepackage[colorlinks=true,citecolor=blue,linkcolor=blue,urlcolor=blue]{hyperref}
\usepackage[usenames]{color}
\usepackage{graphicx}
\usepackage[export]{adjustbox}
\usepackage{subfigure}
\usepackage{amsmath}
\usepackage{epsfig}
\usepackage{dcolumn}
\usepackage{bm}
\usepackage{color}
\usepackage{times}
\usepackage{epstopdf}
\usepackage{amssymb}
\usepackage{amstext}
\usepackage{latexsym}
\usepackage{amsfonts}
\usepackage{psfrag}
\usepackage[utf8]{inputenc}
\usepackage{graphicx}
\usepackage{amsmath}
\usepackage{amssymb}
\usepackage{dsfont}
\usepackage{tensor}
\usepackage{color}
\usepackage{ulem}
\usepackage{upgreek}
\usepackage{braket}
\usepackage{algpseudocode}
\usepackage{algorithm}
\usepackage{orcidlink}  

\UseRawInputEncoding

\begin{document}

\title{Geometric control of maximal entanglement via bound states in the continuum}

\author{Alexis R. Leg\'on \orcidlink{0000-0003-1650-488X}}
\email{legon.oropeza@usm.cl}
\affiliation{Departamento de F\'isica, Universidad T\'ecnica Federico Santa Mar\'ia, Casilla 110 V, Valpara\'iso, Chile}

\author{Mario Miranda Rojas \orcidlink{0000-0002-5449-0565}}
\email{mario.mirandar@usm.cl}
\affiliation{Departamento de F\'isica, Universidad T\'ecnica Federico Santa Mar\'ia, Casilla 110 V, Valpara\'iso, Chile}

\author{Pedro Orellana\orcidlink{0000-0001-7688-4111}}
\email{pedro.orellana@usm.cl}
\affiliation{Departamento de F\'isica, Universidad T\'ecnica Federico Santa Mar\'ia, Casilla 110 V, Valpara\'iso, Chile}

\author{Ariel Norambuena\orcidlink{0000-0001-9496-8765}}
\email{ariel.norambuena@usm.cl}
\affiliation{Departamento de F\'isica, Universidad T\'ecnica Federico Santa Mar\'ia, Casilla 110 V, Valpara\'iso, Chile}

\date{\today}

\begin{abstract}
Bound states in the continuum (BICs) convert dissipative open systems into effectively closed quantum subspaces through destructive interference. We show that two identical giant atoms coupled to a one-dimensional waveguide support BICs that coincide with maximally entangled atomic states. Most importantly, entanglement is predominantly determined by the geometric design; the ratio of intra-atomic connection lengths fixes the concurrence, while the propagation phase between atoms selects a family of Bell-like states. We further analyze the dynamical stability of these maximally entangled BICs under exact time evolution, revealing a clear hierarchy of robustness against parameter perturbations. Our results establish an analytical bridge between symmetry, geometry, entanglement, and BICs in giant-atom waveguide platforms.
\end{abstract}

\maketitle

\textit{Introduction} --- Bound states in the continuum (BICs), first predicted by von Neumann and Wigner~\cite{vN-W} at the dawn of quantum mechanics, are localized states embedded within a continuous spectrum. These states arise from destructive interference that suppresses tunneling, despite their spectral overlap with the continuum. Over the past decade, BICs have been experimentally realized in photonics, acoustics, and related fields, facilitating advanced applications such as high-$Q$ resonances, lasing, and sensing~\cite{Experimentalobservation,ReviewNature,PhysRevA.111.013529}. In quantum systems, BICs are interpreted as dark eigenmodes that form decoherence-free subspaces, making them valuable for preserving coherence and quantum correlations~\cite{Sheremet2023}.

Recent studies in quantum optics have investigated BICs using giant atom devices. These artificial atoms act as quantum emitters that couple to a waveguide at multiple, spatially separated locations~\cite{Facchi2016,Kockum2018,Kannan2020,Dressed.Interference.in.Giant.Superatoms,Legon2025Tunable}. The spatial configuration of these coupling points introduces a nontrivial form factor to the light-matter interaction, producing interference effects absent in point-like emitters. This arrangement enables geometry-dependent control over the emitter's effective coupling to the waveguide, exceeding the capabilities of conventional point-like atoms~\cite{Kockum2018}. Experiments with superconducting giant atoms have shown that waveguide emission can be significantly suppressed while coherent, waveguide-mediated interactions persist~\cite{Kannan2020}. As a result, waveguide quantum electrodynamics (QED) represents a promising approach for realizing long-lived, geometry-protected quantum states.

Quantum entanglement is a fundamental concept in quantum technologies, including computation, communication, and sensing~\cite{Horodecki2009}. Recent research demonstrates that non-interacting giant atoms can achieve long-distance entanglement by employing a waveguide as a shared reservoir~\cite{Yin2023, Weng2024,PhysRevA.111.053711,Xian-Li2025}. Despite these developments, the analytical relationship between bound states in the continuum (BICs) and stationary entanglement remains underexplored. Prior studies in waveguide and cavity quantum electrodynamics (QED) suggest that BICs may correspond to strongly entangled subradiant collective atomic states. Nevertheless, a geometry-programmable analytical mapping to maximally entangled two-qubit states has not yet been established~\cite{FongLaw2017}. This raises the question of whether the geometric configuration of the system can leverage the spectral and interference characteristics of BICs to generate robust, maximally entangled states encoded within giant atoms.

\begin{figure}[ht!]
\centering
\includegraphics[width = \linewidth]{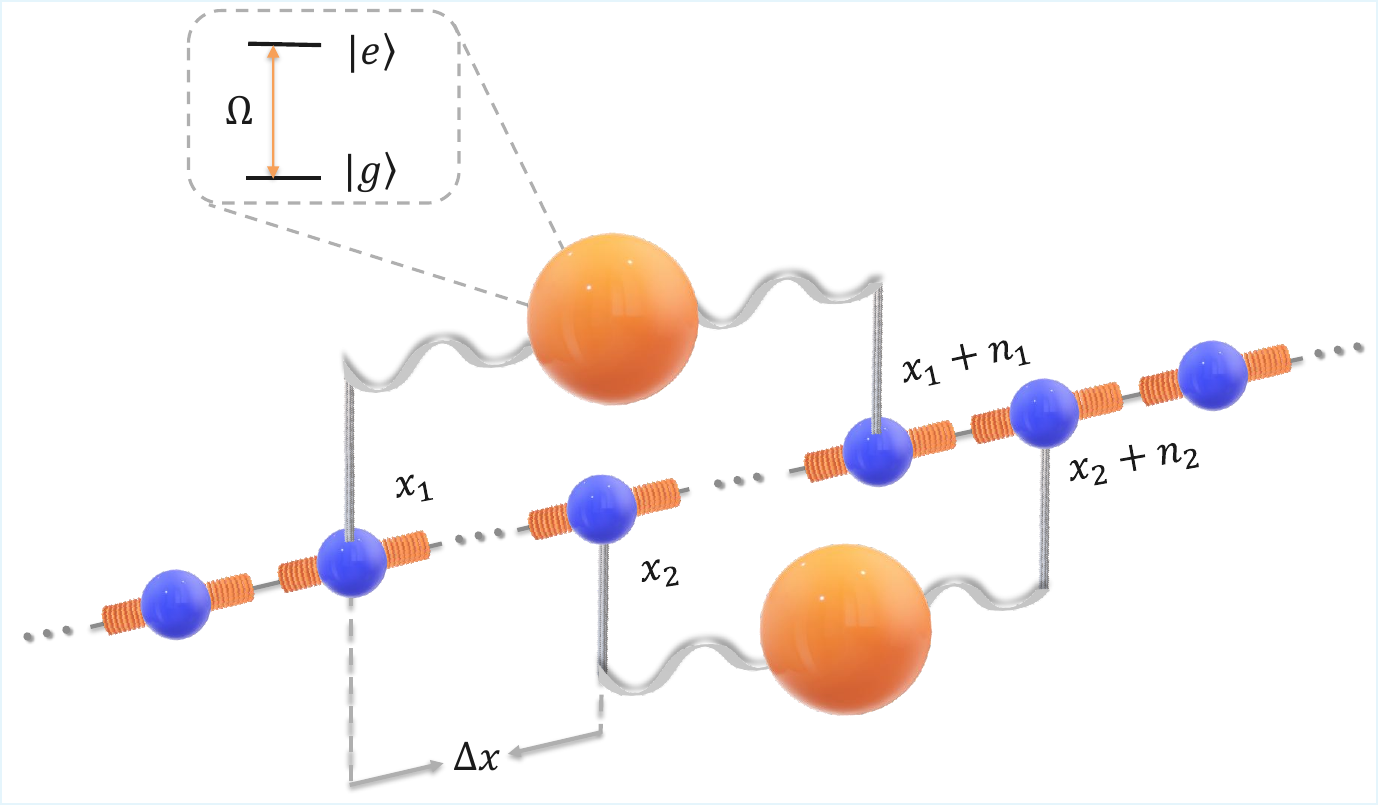}
\caption{
Schematic of two giant atoms coupled to a one-dimensional coupled-resonator waveguide through two connection sites in a braided configuration. Blue spheres represent the waveguide resonators, and orange spheres denote the giant atoms.}
\label{fig:Figure1}
\end{figure}

Geometric control offers an alternative approach for generating entanglement by directing the quantum system along defined spatial trajectories, thereby accumulating phases that affect the quantum state~\cite{Sorensen1999,Sorensen2000}. Representative examples include geometric-phase gates in holonomic schemes~\cite{ZHANG2023} and the control of Bell correlations in Mach-Zehnder interferometers~\cite{Cildiroglu2025}. Nevertheless, the influence of geometric parameters, such as the configuration of coupling points, on the formation of BICs and the stabilization of maximally entangled states has not yet been systematically investigated.

In this Letter, we present a theoretical framework demonstrating that BICs within a two-giant-atom waveguide offer a direct, geometry-based approach to generating robust entanglement. Adjusting the intra-atom length ratio controls the concurrence, while the phase between atoms determines the specific Bell-like state. The present results provide an analytical connection between BIC design and robust, geometry-protected entanglement. Furthermore, analysis of the exact time evolution reveals how the entangled BIC subspace responds to parameter perturbations, demonstrating a hierarchy of robustness consistent with recent BIC-enabled protection mechanisms in giant-atom platforms~\cite{PhysRevA.111.053711,Ingelsten2024}. 

\begin{figure*}[ht!]
\centering
\includegraphics[width = 0.9 \linewidth]{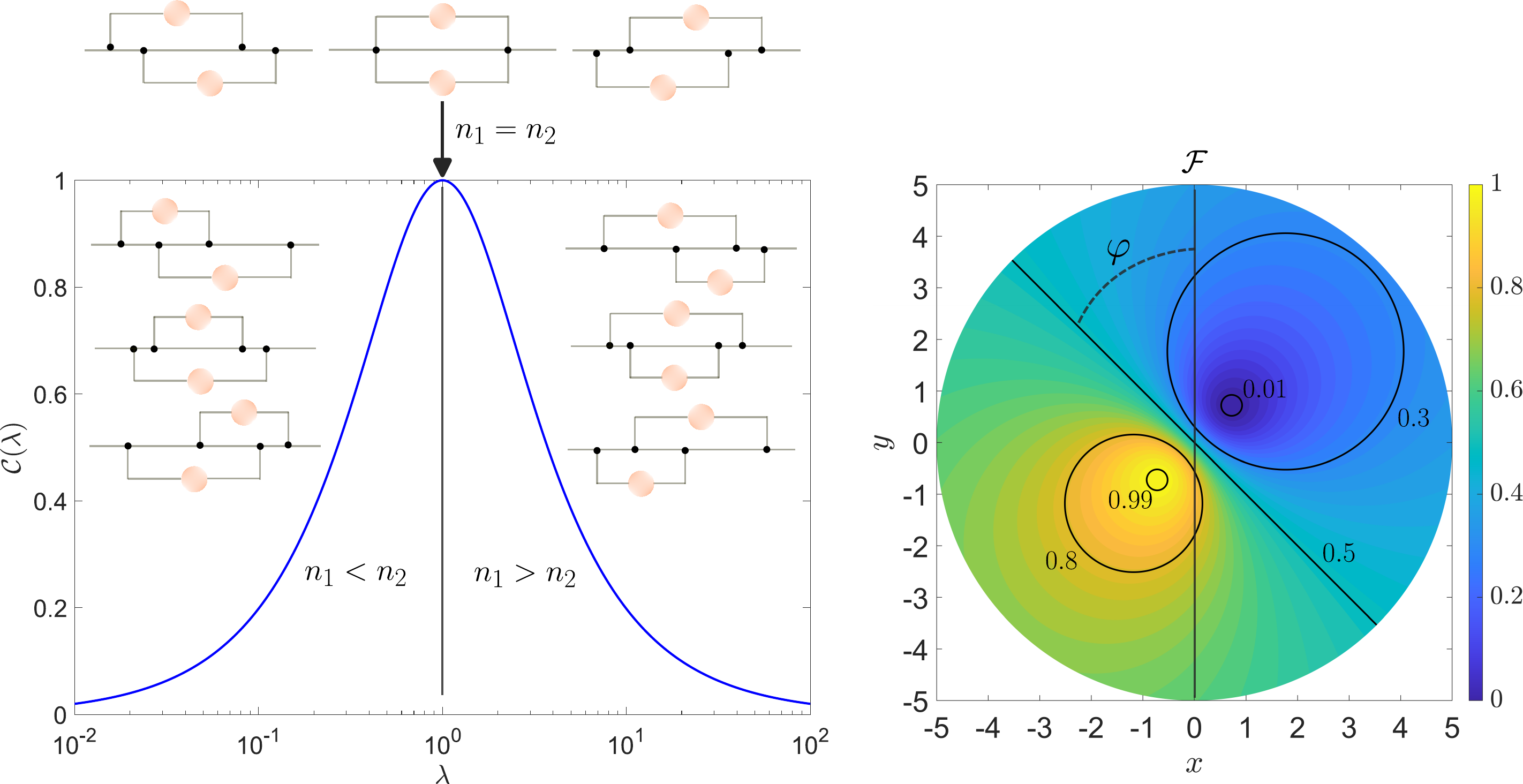}
\caption{(Left) Concurrence $\mathcal{C}(\lambda)$ as a function of the geometric ratio $\lambda = n_1/n_2$ (semi-log scale), characterizing the degree of bipartite entanglement between the two giant atoms. The concurrence is maximal at $\lambda=1$ ($n_1=n_2$). Insets illustrate representative connection geometries across different entanglement regimes. (Right) Polar representation $(x,y)=\lambda(\cos\theta,\sin\theta)$ of the fidelity $\mathcal{F}$ between the atomic BIC state and maximally entangled states $\ket{\Phi}=(\ket{e,g}+ e^{i\varphi}\ket{g,e})/\sqrt{2}$, as a function of $\lambda$ and the relative phase $\theta=k^{\star}\Delta x$. Colors encode $\mathcal{F}$ and solid black curves indicate selected iso-fidelity contours. High-fidelity regions around $\lambda=1$ correspond to maximally entangled BICs.}
\label{fig:Figure2}
\end{figure*}

\textit{Model} --- We consider two identical giant atoms coupled at two spatially separated points each to a one-dimensional coupled-resonator waveguide. In the rotating-wave approximation, the resulting Hamiltonian in momentum space reads~\cite{PhysRevA.111.053711} ($\hbar=1$)
\begin{equation}\label{Hamiltonian}
\begin{aligned}
H &= H_0
+ \sum_{k}\sum_{i=1}^{2}\left(g_{ik}\, \sigma_i^{+}a_k
+ g_{ik}^{\ast}\, \sigma_i^{-}a_k^{\dagger} \right),\\
\omega_k &= \omega_c - 2\xi \cos k, \quad
g_{ik} = \frac{2g}{\sqrt{N_c}}\,A_{k,n_i}\,
e^{-ik\left(x_i+\frac{n_i}{2}\right)},
\end{aligned}
\end{equation}
where $A_{k,n_i} \equiv \cos(k n_i/2)$ is the geometric form factor associated with the intra-atom separation $n_i$ and quasi-momentum $k\in[0,\pi]$. The free Hamiltonian
$H_0=\Omega\sum_{i}\sigma_i^{+}\sigma_i^{-}+\sum_k\omega_k a_k^{\dagger}a_k$
describes the two uncoupled giant atoms and the quantized waveguide, where $\Omega$ is the atomic transition frequency and $\omega_k=\omega_c-2\xi\cos k$ is the waveguide dispersion relation ($\omega_c$ is the bare resonator frequency and $\xi$ is the hopping amplitude). Here, $\sigma_i^{-}=|g\rangle_i\langle e|$ and $a_k$ denote atomic lowering and waveguide annihilation operators, respectively. The waveguide bosonic modes define a discrete band $\omega_k\in[\omega_c-2\xi,\omega_c+2\xi]$, with center at $k=\pi/2$. The coupling amplitudes $g_{ik}$ depend explicitly on the left connection point $x_i$ and the intra-atom separation $n_i$, which encode the system's geometry, see Fig.~\ref{fig:Figure1} for further details. We define $\Delta x\equiv x_2-x_1$ as the distance between the left connection points of the two atoms.

Due to the rotating-wave approximation, the system Hamiltonian~\eqref{Hamiltonian} conserves the total number of excitations, i.e., $[H,N]=0$, with $N=\sum_i \sigma_i^{+}\sigma_i^{-}+\sum_k a_k^{\dagger}a_k$ the excitation number operator. As a result, the Hilbert space decomposes into invariant subspaces labeled by the eigenvalues of $N$. In the single-excitation manifold, the general state reads $\ket{\Psi} = c_1 \ket{e,g,0} +c_2 \ket{g,e,0} +\sum_k \phi_k \ket{g,g,1_k}$, with normalization $|c_1|^2+|c_2|^2+\sum_k|\phi_k|^2=1$. In the case $n_1=n_2$, the system Hamiltonian is also invariant under permutation of the two giant atoms. The symmetric and antisymmetric Bell states $\ket{\Psi_{\pm}}=(\ket{e,g}\pm\ket{g,e})/\sqrt{2}$~\cite{Bell1964} are maximally entangled atomic states, eigenstates of the permutation operator, and compatible with the single-excitation manifold. Bell states are numerically reported in Ref.~\cite{PhysRevA.111.053711}, in the context of giant atoms. However, a broader class of maximally entangled states, $\ket{\Phi(\varphi)}=(\ket{e,g}+e^{i\varphi}\ket{g,e})/\sqrt{2}$ with $\varphi\in[0,2\pi)$~\cite{Nielsen_Chuang_2010}, may arise due to relative phases in the atom-waveguide couplings~\eqref{Hamiltonian}. To quantify relative phases between giant atoms ($g_{2k}/g_{1k}\sim e^{-i\Phi_k}$), we introduce the phase parameter $\Phi_k \equiv k(\Delta x + \Delta n/2)$ between the two atomic excitation pathways, with $\Delta n \equiv n_2 - n_1$.

\textit{BICs and maximally entangled states} --- Starting from the eigenvalue equation $H\ket{\Psi}=E\ket{\Psi}$ within the single-excitation manifold, we find $(E-\Omega)\mathds{1}\mathbf{c}=\boldsymbol{\Sigma}(E)\mathbf{c}$. Here, $\mathds{1}$ is the $2\times2$ identity matrix, $\mathbf{c}=[c_+,c_-]^T$ with $c_{\pm}=(c_1\pm c_2)/\sqrt{2}$ (Bell basis), and $\boldsymbol{\Sigma}(E)$ is a diagonal, energy-dependent matrix describing the self-energy. The elements are $\Sigma_{\pm}(E)=(2g^2/N_c)\sum_k f_k^{\pm}(\Delta x,n_1,n_2)/(E-\omega_k+i\eta)$, where $\eta\to0^{+}$ imposes retarded boundary conditions in the Green’s function. The main interference effect, arising from the geometric properties of the couplings, is expressed by the function
\begin{equation} \label{InterferenceFunction}
  f_k^{\pm}(n_1, n_2, \Delta x) =  \left|A_{k,n_1} \pm A_{k,n_2} e^{i\Phi_k} \right|^2,
\end{equation}
where $\Phi_k$ is the relative phase discussed in the previous section. The above expression makes explicit the geometric origin of the interference: $f_k^{\pm}$ is the squared modulus of the sum of two propagation amplitudes carrying a relative phase $\Phi_k$. The geometry enters through the intra-atom separations $n_i$ and the inter-atom distance $\Delta x$. Writing $\Sigma_{\pm}(E)=\Delta^{\pm}(E)-(i/2)\Gamma^{\pm}(E)$ and using the Sokhotski-Plemelj identity~\cite{Breuer2016}, we obtain the decay rate for each channel $\Gamma^{\pm}(E) = \sum_k (2g^{2}/N_c)/[(E-\omega_k)^{2}+\eta^{2}] f_{k}^{\pm}(\Delta x, n_1, n_2)$. The Lamb shift is given by $\Delta^{\pm}(E)=\sum_k (2g^{2}/N_c)(E-\omega_k)/[(E-\omega_k)^{2}+\eta^{2}] f_{k}^{\pm}(\Delta x, n_1, n_2)$. Note that both quantities, $\Gamma^{\pm}(E)$ and $\Delta^{\pm}(E)$, are weighted by the interference structure $f_k^{\pm}$~\eqref{InterferenceFunction}.

A BIC corresponds to an in-band eigenenergy $E_{\rm BIC}$ for which the radiative channel vanishes~\cite{Friedrich1985}, i.e., $\Gamma^{\pm}(E_{\rm BIC})=0$. A particularly robust route is provided by the giant-atom form factor condition $A_{k,n_i}=0$, which enforces $\cos(k^\star n_i/2)=0$ at the resonant wave vector $k^\star$, and hence suppresses radiation independently of $\Delta x$. Thus, the robust condition to have a BIC is defined by 
\begin{equation}\label{robust_BIC_condition}
k^{\star} n_i=(2\ell_{i}+1)\pi,\qquad \ell_{i}\in\mathbb{Z}.
\end{equation}
More generally, BIC formation can be stated as an on-shell cancellation of the radiative amplitude. In the original atomic basis, this reads $c_1 g_{1k}^{\ast}+c_2 g_{2k}^{\ast}=0$ at the resonant wave vectors $k=\pm k^\star$. Combining this condition with the robust BIC constraint yields the amplitude relation $c_2=-c_1 (n_1/n_2)\,e^{-ik^\star \Delta x}$. The resulting stationary BIC state is parametrized as
\begin{eqnarray}\label{BIC}
\ket{\Psi_{\text{BIC}}}
&=& \frac{\ket{e,g}-\lambda\, e^{-i\phi_{k^{\star}}}\ket{g,e}}{\sqrt{1+\lambda^2}}
\otimes \ket{0} = \ket{\Psi_{\rm ga}^{\rm BIC}}\otimes \ket{0},
\end{eqnarray}
where $\lambda = n_1/n_2$ is the ratio of intra-atom connection lengths, $\phi_{k^{\star}} = k^{\star} \Delta x$ is a propagation phase, and $\ket{\Psi_{\rm ga}^{\rm BIC}}$ is the atomic wave function. The density matrix is given by a product $\rho_{\rm BIC}=\rho_{\rm ga}^{\rm BIC}\otimes \ket{0}\bra{0}$, with $\rho_{\rm ga}^{\rm BIC}= \ket{\Psi_{\rm ga}^{\rm BIC}} \bra{\Psi_{\rm ga}^{\rm BIC}}$ the pure atomic state. The degree of bipartite entanglement can be quantified by the concurrence introduced by Wootters~\cite{Wootters1998}. For the giant atom density matrix, the concurrence can be computed as $\mathcal{C} = \max\left\{0, \lambda_1 - \lambda_2 - \lambda_3 - \lambda_4 \right\}$, where $\lambda_i$ are the eigenvalues, in decreasing order ($\lambda_1 \geq \lambda_2 \geq \lambda_3 \geq \lambda_4$), of the matrix $R = (\rho^{1/2} \widetilde{\rho} \rho^{1/2})^{1/2}$, with $\widetilde{\rho} = \left( \sigma_y \otimes \sigma_y \right)
\rho^{\ast} \left( \sigma_y \otimes \sigma_y \right)$ and $\rho = \rho_{\rm ga}^{\rm BIC}$. The exact concurrence for the giant atoms reads
\begin{equation}
\mathcal{C}(\lambda)=\frac{2|\lambda|}{1+\lambda^2},
\end{equation}
which is entirely geometric, as it depends only on the ratio $\lambda \equiv n_1/n_2$. Importantly, the scaling property $\mathcal{C}(1/\lambda)=\mathcal{C}(\lambda)$ follows directly from the permutation symmetry: $n_1 \rightarrow n_2$ and $n_2 \rightarrow n_1$, which maps $\lambda \to 1/\lambda$. The plot of the concurrence in logarithmic scale is shown in the left panel of Fig.~\ref{fig:Figure2}, illustrating the connection between geometry and entanglement for the BIC in different giant atom configurations. The fidelity between a BIC and the maximally entangled states $|\Phi(\varphi)\rangle=(\ket{e,g}+ e^{i\varphi}\ket{g,e})/\sqrt{2}$ is given by
\begin{eqnarray}
\mathcal{F} = \frac{1 -\mathcal{C}(\lambda)\cos(k^\star \Delta x-\varphi)}{2},
\end{eqnarray}
where $\mathcal{F}=|\langle \Psi_{\rm ga}^{\rm BIC}|\Phi(\varphi)\rangle|^2$ for pure states. The concurrence $\mathcal{C}$, and hence the geometry factor $\lambda$, sets the bounds $(1-\mathcal{C})/2\leq \mathcal{F}\leq(1+\mathcal{C})/2$. Maximally entangled BIC states are obtained for $n_1 = n_2$ and $k^{\star}\Delta x = \varphi + 2m\pi$, with $m\in\mathbb{Z}$. Thanks to the phase degree of freedom $k^{\star}\Delta x$ given in Eq.~\eqref{BIC}, the BIC state can parametrize different entangled states with a relative phase. The symmetry behavior of the fidelity is summarized by the polar map in the right panel of Fig.~\ref{fig:Figure2}, where the radial coordinate $\lambda$ fixes the concurrence and the angular coordinate $k^\star\Delta x$ determines the Bell-like sector. In particular, different values of $\varphi$ introduce a rotational symmetry on the fidelity map.

\begin{figure*}[ht!]
\centering
\includegraphics[width = 0.9 \linewidth]{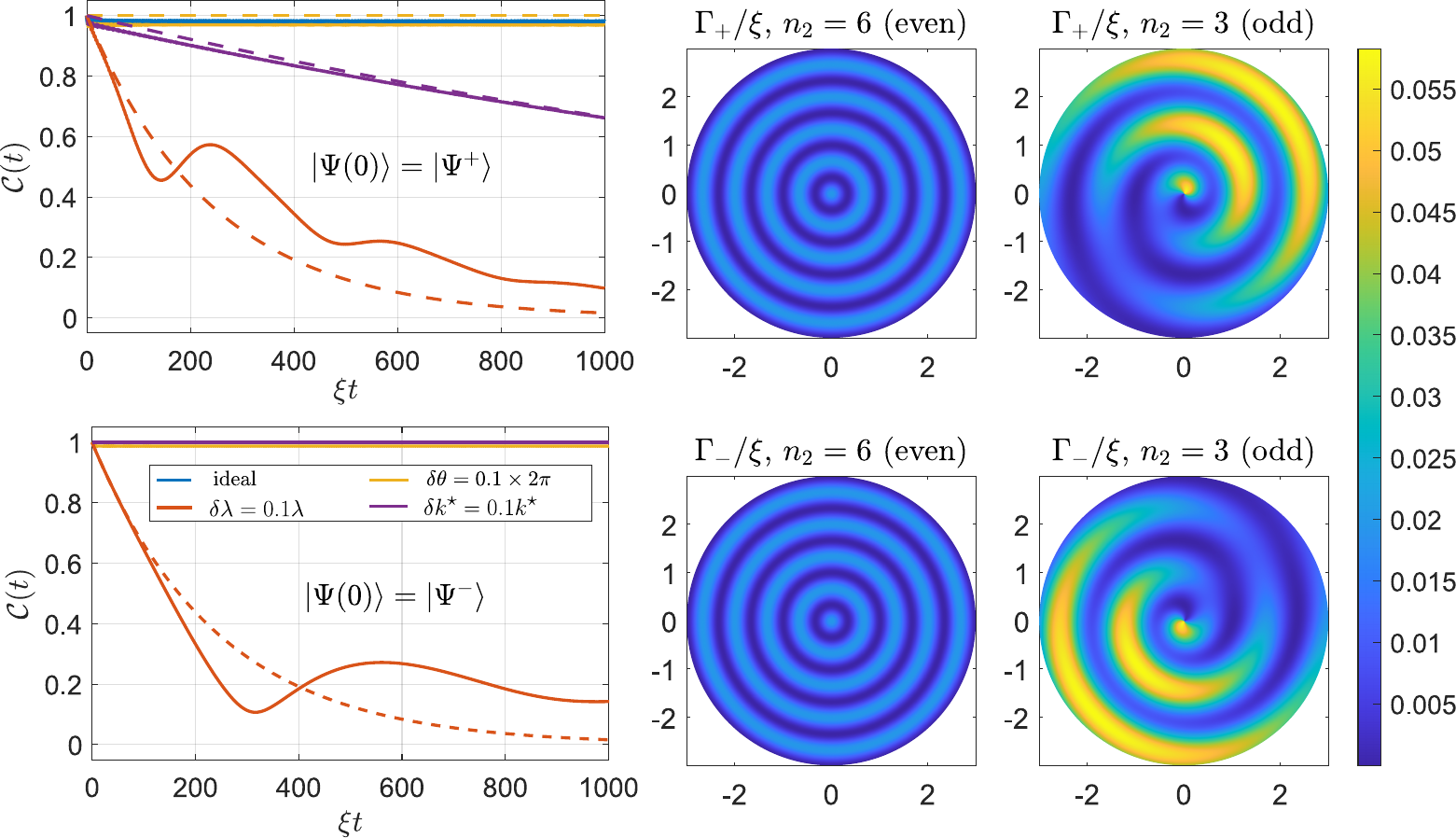}
\caption{(Left) Time evolution of the concurrence $C(t)$ for initial atomic states aligned with BICs in the symmetric (top) and antisymmetric (bottom) Bell states, $\ket{\Psi^{\pm}}$. The ideal geometric configuration $\lambda = 1$ and $k^{\star} = \pi/2$ (solid blue) yields long-lived entanglement, while evolution with $10\%$ detunings in the parameters $\lambda$, $\theta = k^{\star} \Delta x$, and $k^{\star}$ shows deviations from the ideal case. Dashed curves correspond to the Markovian approximation $C_{\pm}(t)=C_{\pm}(0)e^{-\Gamma^{\pm}(\Omega)t}$, where decay rates $\Gamma^{\pm}(\Omega)$ are given in Eq.~\eqref{Rate}. (Right) Polar heat maps of the normalized decay rates $\Gamma^{\pm}/\xi$ as a function of the geometric coordinates $(x,y)=\lambda(\cos\theta,\sin\theta)$. For even $n_2$, the vanishing of the giant-atom form factor at $k^{\star}$ suppresses angular interference, yielding purely concentric decay rings, whereas for odd $n_2$ residual interference produces strongly anisotropic decay landscapes. Simulation parameters: $\xi=1$, $g=0.1\xi$, $\Omega=\omega_c$. The waveguide is discretized with a large number of modes ($N_c = 4 \times 501$) to symmetrically resolve the resonant momenta $k = \pm k^\star$ ($k^{\star} = \pi/2$), ensuring convergence of the exact diagonalization.}
\label{fig:Figure3}
\end{figure*}

\textit{Entanglement dynamics} --- Previous research on dynamical entanglement in giant atoms has primarily examined non-Markovian effects in the presence of gauge fields~\cite{photonics12060612} and Markovian dynamics using effective Hamiltonians within the weak-coupling regime~\cite{PhysRevA.111.053711}. Additionally, Laplace transform techniques have been employed to study strong coupling~\cite{wu2025}, including investigations of long-range tunneling in ultra-cold atoms~\cite{Yang2025}. However, a comprehensive framework that integrates exact non-Markovian dynamics, BICs, and the geometric robustness of maximally entangled BIC states has yet to be established. To elucidate the dynamic correspondence between BICs and maximally entangled atomic states, the evolution is analyzed in the symmetric and antisymmetric Bell basis. This basis offers a natural framework for capturing the relevant dynamical features. When restricted to the single-excitation manifold, the state is given by
$
\ket{\Psi(t)} = c_{+}(t)\ket{\Psi^{+},0} + c_{-}(t)\ket{\Psi^{-},0} + \sum_k c_k(t)\ket{g,g,1_k},
$
where $c_{\pm}(t)$ denote the Bell-channel amplitudes. In the interaction picture with respect to the bare Hamiltonian $H_0$, the two Bell components evolve independently and couple to the waveguide through the effective coupling
$W_k^{\pm}=(g_{1k}\pm g_{2k})/\sqrt{2}$. Eliminating the bosonic degrees of freedom for an initially empty waveguide yields the integro-differential equations
\begin{equation}
\dot c_\pm(t)= -\int_0^t K_\pm(\tau) c_\pm(t-\tau) d\tau,
\label{EntanglementDynamics}
\end{equation}
with memory kernels $K_\pm(\tau)=\sum_k |W_k^\pm|^2 e^{i(\Omega-\omega_k)\tau}$ encode waveguide-induced interference and non-Markovian effects. Applying the Laplace transform to Eq.~\eqref{EntanglementDynamics} yields the time-dependent solution $c_\pm(t)= (2\pi i)^{-1}\int_{\gamma-i\infty}^{\gamma+i\infty}e^{st}c_\pm(0)/[s+K_\pm(s)]ds$, where the Bromwich contour is positioned to the right of all singularities. According to the residue theorem, the system dynamics are determined by the poles of $s+K_\pm(s)$. A BIC corresponds to a real in-band solution of $E_{\rm BIC}-\Omega-\Sigma_\pm(E_{\rm BIC})=0$ with a vanishing decay rate, $\Gamma_\pm(E_{\rm BIC})=0$. This Laplace pole is located on the imaginary axis of the complex $s$-plane, $s = -i(E_{\rm BIC}-\Omega)$, resulting in a Bell state that does not decay. 

Figure~\ref{fig:Figure3} illustrates the dynamical properties and geometric influences on entanglement and damping rates. Atoms are initialized in maximally entangled BICs, and exact numerical simulations are performed using the full Hamiltonian and symmetries, where the concurrence is computed as $C(t) = \left|\,|c_+(t)|^2-|c_-(t)|^2
+2i\,\mathrm{Im}\!\left[c_-(t)c_+^*(t)\right]\right|$. The ideal geometric configuration yields long-lived concurrence ($C(t) = 1$), while controlled 10\% detuning tests on the ideal parameters $(\lambda = 1, \theta = k^{\star} \Delta x, k^{\star} = \pi/2)$ demonstrate the system's robustness. We observe a clear hierarchy: variations in the geometric ratio $\lambda$ are the primary source of entanglement degradation ($C(t)<0.4$ for $\xi t = 400$), whereas similar detunings in the propagation phase $\theta$ or in the resonant wave vector $k^{\star}$ have only a minor impact on $C(t)$ in the symmetric channel. This behavior reflects the channel-dependent interference encoded in $W_{k^{\star}}^{\pm}$. For short times ($\xi t \leq 100$), the dynamics are well described by the Markovian approximation $C_{\pm}(t) \approx C_{\pm}(0) \mbox{exp}(-\Gamma^{\pm}(\Omega)t)$, with decay rates
\begin{equation}
\Gamma^{\pm}(\Omega)=\frac{2g^2}{\xi|\sin k(\Omega)|}
f^{\pm}_{k(\Omega)}(\Delta x,n_1,n_2),
\label{Rate}
\end{equation}
where $k(E)=\cos^{-1}[(\omega_c-\Omega)/(2\xi)]$.
Equation~\eqref{Rate} establishes a direct quantitative link between entanglement decay and geometry. The geometric dependence is made explicit by the polar maps of $\Gamma^{\pm}/\xi$ in Fig.~\ref{fig:Figure3} (right panel), shown as functions of the coordinates $(x,y)=\lambda(\cos\theta,\sin\theta)$. Fixing $k^{\star}=\pi/2$ reveals a parity effect in $n_2$, for even $n_2$, the giant-atom form factor vanishes, $A_{k^{\star},n_2} = \cos(\pi n_2/4)$, suppressing angular interference between $A_{k^{\star},n_1}$ and $A_{k^{\star},n_2}$, producing purely radial landscapes with maxima and minima. In contrast, for odd $n_2$, residual interference leads to strongly anisotropic patterns arising from the phase $\Phi_{k^{\star}}$. Together, the exact concurrence dynamics and rate maps demonstrate that geometry not only fixes the amount of entanglement, but also selects which Bell channel remains protected throughout the evolution.

\textit{Conclusions} --- We have shown that BICs in a two-giant-atom waveguide system provide a simple and powerful geometric route to engineering maximally entangled states. Interference from multiple coupling points yields a continuous family of pure two-qubit states in the single-excitation manifold. These states are exactly decoupled from the photonic continuum. Our analysis reveals a clear geometric design principle. The ratio of the intra-atom connection lengths, $\lambda = n_1/n_2$, uniquely fixes the concurrence, while the phase $k^\star\Delta x$ acts as a Bell selector determining the fidelity. When the connection lengths are balanced, $n_1=n_2$, maximally entangled states emerge as BICs without requiring dynamical control or engineered dissipation.

Beyond their stationary character, we demonstrate that these states exhibit strong dynamical robustness under parameter perturbations. Tracking their exact evolution shows distinct levels of stability in responses to variations in geometric parameters. The states also interact preferentially with specific Bell channels (routes for entangled quantum information), revealing a clear link between their isolated energy behaviors, their physical structure, and how well they retain their properties over time. An interesting direction is to use optimal quantum control to generate and protect these geometric maximally entangled BICs for two or more giant atoms. Since the underlying mechanism relies solely on waveguide-mediated interference, our results are directly applicable to current superconducting-circuit and photonic implementations of giant atoms, positioning BICs as a versatile geometric resource for robust entanglement engineering in waveguide QED and related open quantum systems.

\section{Acknowledgments}
A.R.L. and M.M. acknowledge the financial support from the project Fondecyt No. 3240726. P.A.O. acknowledges support from DGIIE USM PI-LIR-24-10 and FONDECYT Grants No. 122070, 1230933. A.N. acknowledges the financial support from the project Fondecyt Regular No. 1251131 and ANID Anillo Project ATE250066.

\bibliography{Refs}

@article{vN-W,
  title = {Bound states in the continuum},
  author = {J. von Neumann and E. Wigner},
  journal = {Z. Phys. Chem., Abt. A},
  volume = {30},
  pages = {465},
  year = {1929},
}

@article{ReviewNature,
author = {Hsu, Chia Wei and Zhen, Bo and Stone, A Douglas and Joannopoulos, John D and Solja{\v{c}}i{\'{c}}, Marin},
journal = {Nature Reviews Materials},
month = {jul},
pages = {16048},
publisher = {Macmillan Publishers Limited},
title = {{Bound states in the continuum}},
url = {http://dx.doi.org/10.1038/natrevmats.2016.48 http://10.0.4.14/natrevmats.2016.48},
volume = {1},
year = {2016}
}

@article{Experimentalobservation,
  title = {Experimental Observation of Optical Bound States in the Continuum},
  author = {Plotnik, Yonatan and Peleg, Or and Dreisow, Felix and Heinrich, Matthias and Nolte, Stefan and Szameit, Alexander and Segev, Mordechai},
  journal = {Phys. Rev. Lett.},
  volume = {107},
  issue = {18},
  pages = {183901},
  numpages = {4},
  year = {2011},
  month = {Oct},
  publisher = {American Physical Society},
  doi = {10.1103/PhysRevLett.107.183901},
}

@article{PhysRevA.111.013529,
  title = {Bound states in the continuum in a double whispering-gallery resonator},
  author = {Leg\'on, Alexis R. and Ahumada, M. and Ramos-Andrade, J. P. and Molina, Rafael A. and Orellana, P. A.},
  journal = {Phys. Rev. A},
  volume = {111},
  issue = {1},
  pages = {013529},
  numpages = {13},
  year = {2025},
  month = {Jan},
  publisher = {American Physical Society},
  doi = {10.1103/PhysRevA.111.013529},
  url = {https://link.aps.org/doi/10.1103/PhysRevA.111.013529}
}

@article{PhysRevA.111.053711,
  title = {High-fidelity generation of Bell and $W$ states in a giant-atom system via bound states in the continuum},
  author = {Weng, Mingzhu and Yu, Hongwei and Wang, Zhihai},
  journal = {Phys. Rev. A},
  volume = {111},
  issue = {5},
  pages = {053711},
  numpages = {7},
  year = {2025},
  month = {May},
  publisher = {American Physical Society},
  doi = {10.1103/PhysRevA.111.053711},
  url = {https://link.aps.org/doi/10.1103/PhysRevA.111.053711}
}

@article{Sheremet2023,
  author = {Sheremet, A. S. and Petrov, M. I. and Iorsh, I. V. and Poshakinskiy, A. V. and Poddubny, A. N.},
  title = {Waveguide quantum electrodynamics: Collective radiance and photon-photon correlations},
  journal = {Rev. Mod. Phys.},
  volume = {95},
  pages = {015002},
  year = {2023},
  doi = {10.1103/RevModPhys.95.015002}
}

@article{Facchi2016,
  author = {Facchi, P. and Lombardo, F. and Pascazio, S. and Pepe, F. V. and Yonac, M.},
  title = {Bound states in the continuum for an array of quantum emitters},
  journal = {Phys. Rev. A},
  volume = {94},
  pages = {043839},
  year = {2016},
  doi = {10.1103/PhysRevA.94.043839}
}

@article{Kockum2018,
  author = {Kockum, A. F. and Johansson, G. and Nori, F.},
  title = {Decoherence-free interaction between giant atoms in waveguide quantum electrodynamics},
  journal = {Phys. Rev. Lett.},
  volume = {120},
  pages = {140404},
  year = {2018},
  doi = {10.1103/PhysRevLett.120.140404}
}

@article{Legon2025Tunable,
  author  = {Leg\'{o}n, Alexis R. and Miranda, Mario and Orellana, P. A.},
  title   = {Tunable quantum photonic routing using a coupled giant-atom-like array},
  journal = {arXiv},
  volume  = {2511.13992},
  year    = {2025},
  note    = {arXiv:2511.13992 [quant-ph]},
  url     = {https://arxiv.org/abs/2511.13992}
}

@article{Kannan2020,
  title = {Waveguide quantum electrodynamics with superconducting artificial giant atoms},
  author = {Kannan, Bharath and Ruckriegel, Max J. and Campbell, Daniel L. and Frisk Kockum, Anton and Braumüller, Jochen and Kim, David K. and Kjaergaard, Morten and Krantz, Philip and Melville, Alexander and Niedzielski, Bethany M. and Vepsäläinen, Antti and Winik, Roni and Yoder, Jonilyn L. and Nori, Franco and Orlando, Terry P. and Gustavsson, Simon and Oliver, William D.},
  journal = {Nature},
  volume = {583},
  issue = {7818},
  pages = {775--779},
  year = {2020},
  month = {Jul},
  doi = {10.1038/s41586-020-2529-9},
  url = {https://doi.org/10.1038/s41586-020-2529-9}
}

@article{Dressed.Interference.in.Giant.Superatoms,
  title = {Dressed Interference in Giant Superatoms: Entanglement Generation and Transfer},
  author = {Du, Lei and Wang, Xin and Kockum, Anton Frisk and Splettstoesser, Janine},
  journal = {Phys. Rev. Lett.},
  volume = {135},
  issue = {22},
  pages = {223601},
  numpages = {9},
  year = {2025},
  month = {Nov},
  publisher = {American Physical Society},
  doi = {10.1103/crzs-k718},
  url = {https://link.aps.org/doi/10.1103/crzs-k718}
}

@article{Wootters1998,
  title = {Entanglement of Formation of an Arbitrary State of Two Qubits},
  author = {Wootters, William K.},
  journal = {Phys. Rev. Lett.},
  volume = {80},
  issue = {10},
  pages = {2245--2248},
  numpages = {0},
  year = {1998},
  month = {Mar},
  publisher = {American Physical Society},
  doi = {10.1103/PhysRevLett.80.2245},
  url = {https://link.aps.org/doi/10.1103/PhysRevLett.80.2245}
}

@article{Horodecki2009,
  title = {Quantum entanglement},
  author = {Horodecki, Ryszard and Horodecki, Pawe\l{} and Horodecki, Micha\l{} and Horodecki, Karol},
  journal = {Rev. Mod. Phys.},
  volume = {81},
  issue = {2},
  pages = {865--942},
  numpages = {0},
  year = {2009},
  month = {Jun},
  publisher = {American Physical Society},
  doi = {10.1103/RevModPhys.81.865},
  url = {https://link.aps.org/doi/10.1103/RevModPhys.81.865}
}

@article{Yin2023,
  title = {Generation of two-giant-atom entanglement in waveguide-QED systems},
  author = {Yin, Xian-Li and Liao, Jie-Qiao},
  journal = {Phys. Rev. A},
  volume = {108},
  issue = {2},
  pages = {023728},
  numpages = {13},
  year = {2023},
  month = {Aug},
  publisher = {American Physical Society},
  doi = {10.1103/PhysRevA.108.023728},
  url = {https://link.aps.org/doi/10.1103/PhysRevA.108.023728}
}

@article{Weng2024,
  title = {Interaction and entanglement engineering in a driven-giant-atom setup with a coupled resonator waveguide},
  author = {Weng, Mingzhu and Wang, Xin and Wang, Zhihai},
  journal = {Phys. Rev. A},
  volume = {110},
  issue = {2},
  pages = {023721},
  numpages = {12},
  year = {2024},
  month = {Aug},
  publisher = {American Physical Society},
  doi = {10.1103/PhysRevA.110.023721},
  url = {https://link.aps.org/doi/10.1103/PhysRevA.110.023721}
}

@article{Xian-Li2025,
  title = {Giant-atom dephasing dynamics and entanglement generation in a squeezed vacuum reservoir},
  author = {Yin, Xian-Li and Lee, Heung-wing Joseph and Zhang, Guofeng},
  journal = {Phys. Rev. A},
  volume = {111},
  issue = {3},
  pages = {033707},
  numpages = {14},
  year = {2025},
  month = {Mar},
  publisher = {American Physical Society},
  doi = {10.1103/PhysRevA.111.033707},
  url = {https://link.aps.org/doi/10.1103/PhysRevA.111.033707}
}

@Article{photonics12060612,
AUTHOR = {Yannopapas, Vassilios},
TITLE = {Entanglement Dynamics of Two Giant Atoms Embedded in a One-Dimensional Photonic Lattice with a Synthetic Gauge Field},
JOURNAL = {Photonics},
VOLUME = {12},
YEAR = {2025},
NUMBER = {6},
ARTICLE-NUMBER = {612},
URL = {https://www.mdpi.com/2304-6732/12/6/612},
ISSN = {2304-6732},
DOI = {10.3390/photonics12060612}
}

@misc{wu2025,
      title={Strongly coupled giant-atom waveguide quantum electrodynamics}, 
      author={Zong-Wei Wu and Jun-Hong An},
      year={2025},
      eprint={2511.01300},
      archivePrefix={arXiv},
      primaryClass={quant-ph},
      url={https://arxiv.org/abs/2511.01300}, 
}

@article{Yang2025,
author={Yang, Yuan-Xing
and Bai, Si-Yuan
and An, Jun-Hong},
title={Long-range quantum tunneling via matter waves},
journal={Communications Physics},
year={2025},
month={Jan},
day={02},
volume={8},
number={1},
pages={5},
issn={2399-3650},
doi={10.1038/s42005-024-01924-y},
url={https://doi.org/10.1038/s42005-024-01924-y}
}

@article{Sorensen1999,
  title = {Quantum Computation with Ions in Thermal Motion},
  author = {S\o{}rensen, Anders and M\o{}lmer, Klaus},
  journal = {Phys. Rev. Lett.},
  volume = {82},
  issue = {9},
  pages = {1971--1974},
  numpages = {0},
  year = {1999},
  month = {Mar},
  publisher = {American Physical Society},
  doi = {10.1103/PhysRevLett.82.1971},
  url = {https://link.aps.org/doi/10.1103/PhysRevLett.82.1971}
}

@article{Sorensen2000,
  title = {Entanglement and quantum computation with ions in thermal motion},
  author = {S\o{}rensen, Anders and M\o{}lmer, Klaus},
  journal = {Phys. Rev. A},
  volume = {62},
  issue = {2},
  pages = {022311},
  numpages = {11},
  year = {2000},
  month = {Jul},
  publisher = {American Physical Society},
  doi = {10.1103/PhysRevA.62.022311},
  url = {https://link.aps.org/doi/10.1103/PhysRevA.62.022311}
}

@article{ZHANG2023,
title = {Geometric and holonomic quantum computation},
journal = {Physics Reports},
volume = {1027},
pages = {1-53},
year = {2023},
note = {Geometric and holonomic quantum computation},
issn = {0370-1573},
doi = {https://doi.org/10.1016/j.physrep.2023.07.004},
url = {https://www.sciencedirect.com/science/article/pii/S0370157323002065},
author = {Jiang Zhang and Thi Ha Kyaw and Stefan Filipp and Leong-Chuan Kwek and Erik Sjöqvist and Dianmin Tong}
}

@article{Cildiroglu2025,
  title = {Geometric control of Bell correlations in path-entangled systems},
  author = {Cildiroglu, H. O.},
  journal = {Phys. Rev. A},
  volume = {112},
  issue = {6},
  pages = {062231},
  numpages = {6},
  year = {2025},
  month = {Dec},
  publisher = {American Physical Society},
  doi = {10.1103/bcsp-2w7b},
  url = {https://link.aps.org/doi/10.1103/bcsp-2w7b}
}

@article{Breuer2016,
  title = {Colloquium: Non-Markovian dynamics in open quantum systems},
  author = {Breuer, H. and Laine, E. and Piilo, J. and Vacchini, B.},
  journal = {Rev. Mod. Phys.},
  volume = {88},
  issue = {2},
  pages = {021002},
  numpages = {24},
  year = {2016},
  publisher = {American Physical Society},
  doi = {10.1103/RevModPhys.88.021002},
}

@article{Bell1964,
  title = {On the Einstein Podolsky Rosen paradox},
  author = {Bell, J. S.},
  journal = {Physics Physique Fizika},
  volume = {1},
  issue = {3},
  pages = {195--200},
  numpages = {6},
  year = {1964},
  month = {Nov},
  publisher = {American Physical Society},
  doi = {10.1103/PhysicsPhysiqueFizika.1.195},
  url = {https://link.aps.org/doi/10.1103/PhysicsPhysiqueFizika.1.195}
}

@article{Friedrich1985,
  title = {Interfering resonances and bound states in the continuum},
  author = {Friedrich, H. and Wintgen, D.},
  journal = {Phys. Rev. A},
  volume = {32},
  issue = {6},
  pages = {3231--3242},
  numpages = {0},
  year = {1985},
  month = {Dec},
  publisher = {American Physical Society},
  doi = {10.1103/PhysRevA.32.3231},
  url = {https://link.aps.org/doi/10.1103/PhysRevA.32.3231}
}

@book{Nielsen_Chuang_2010, 
place={Cambridge}, 
title={Quantum Computation and Quantum Information}, 
publisher={Cambridge University Press}, 
author={Nielsen, Michael A. and Chuang, Isaac L.}, 
year={2010}
}

@article{FongLaw2017,
  title = {Bound state in the continuum by spatially separated ensembles of atoms in a coupled-cavity array},
  author = {Fong, P. T. and Law, C. K.},
  journal = {Phys. Rev. A},
  volume = {96},
  issue = {2},
  pages = {023842},
  numpages = {8},
  year = {2017},
  month = {Aug},
  publisher = {American Physical Society},
  doi = {10.1103/PhysRevA.96.023842},
  url = {https://link.aps.org/doi/10.1103/PhysRevA.96.023842}
}

@article{Ingelsten2024,
  title = {Avoiding decoherence with giant atoms in a two-dimensional structured environment},
  author = {Raaholt Ingelsten, Emil and Kockum, Anton Frisk and Soro, Ariadna},
  journal = {Phys. Rev. Res.},
  volume = {6},
  issue = {4},
  pages = {043222},
  numpages = {20},
  year = {2024},
  month = {Dec},
  publisher = {American Physical Society},
  doi = {10.1103/PhysRevResearch.6.043222},
  url = {https://link.aps.org/doi/10.1103/PhysRevResearch.6.043222}
}

\end{document}